\documentclass[osajnl,twocolumn,showpacs]{revtex4}  

\usepackage{graphicx}
\usepackage{simon}

\begin{document}

\title{Dynamic effects in nonlinear magneto-optics of atoms and molecules}
\author{E. B. Alexandrov}
\email{ealex@online.ru} \affiliation{A. F. Ioffe Physico-Technical
Institute, 26 Polytekhnicheskaya, St Petersburg 194021, Russian
Federation}
\author{M. Auzinsh}
\email{mauzins@latnet.lv} \affiliation{Department of Physics,
University of Latvia,
 19 Rainis blvd, Riga, LV-1586, Latvia}
\author{D. Budker} \email{budker@socrates.berkeley.edu}
\affiliation{Department of Physics, University of California,
Berkeley, CA 94720-7300} \affiliation{Nuclear Science Division,
Lawrence Berkeley National Laboratory, Berkeley CA 94720}
\author{D. F. Kimball}
\affiliation{Department of Physics, University of California,
Berkeley, CA 94720-7300}
\author{S. M. Rochester}
\affiliation{Department of Physics, University of California,
Berkeley, CA 94720-7300}
\author{V. V. Yashchuk}
\affiliation{Advanced Light Source Division, Lawrence Berkeley
National Laboratory, Berkeley CA 94720}

\date{\today}


\begin{abstract}
A brief review is given of topics relating to dynamical processes
arising in nonlinear interactions between light and resonant
systems (atoms or molecules) in the presence of a magnetic field.
\end{abstract}

\maketitle
\tableofcontents

\section{Introduction}

The study of resonant nonlinear magneto-optical effects (NMOE) is
an active area of research with a rich history going back to the
early work on optical pumping. These effects result from the
resonant interaction of light with an atomic or molecular system
in the presence of additional external electromagnetic fields
(most commonly a uniform magnetic field, but we will also consider
electric fields). A recent review by Budker \textit{et
al.}\cite{Bud2002RMP} deals primarily with the effects in atoms;
the molecular case is detailed in a monograph by Auzinsh and
Ferber.\cite{Auz95} In this paper, we focus on the dynamical
aspects of such interactions, attempting to offer the reader a
unified view on various diverse phenomena and techniques while
avoiding excessive repetition of material already discussed in the
previous reviews. Interest in the application of dynamic NMOE to
experimental techniques has been increasing in recent years. The
use of dynamic effects allows one to do things difficult with
steady-state NMOE, for example, high-sensitivity Earth-field-range
magnetometry. Dynamic effects can also allow information (such as
the value of energy-level splittings) that would normally be
obtained from high-resolution spectroscopy to be extracted from
direct measurements of frequency, a more robust technique.

Nonlinear magneto-optical effects (specifically those related to
coherence phenomena\cite{Bud2002RMP}) are observed when a
optically polarized medium undergoes quantum beats (Sec.\
\ref{Sec:QB}) under the influence of an external field, and so
influences the polarization and/or intensity of a transmitted
probe light beam. While this is inherently a dynamical process on
the microscopic level, a macroscopic ensemble of
particles\cite{Note:Particle} reaches a steady state in about the
polarization relaxation time if the external parameters are held
constant. Thus, in order to observe quantum beat dynamics, these
parameters must be varied at a rate significant compared to the
polarization relaxation rate. One approach, discussed in Sec.\
\ref{Sec:PulsedPump}, is to produce polarization with a pulse of
pump light, and then observe the effect of the subsequent quantum
beat dynamics on the optical properties of the medium.
Alternatively, the method of \emph{beat resonances} (Sec.\
\ref{Sec:BeatRes}) can be used, in which an experimental parameter
[the amplitude (Sec.\ \ref{Sec:SynchPump}), frequency (Sec.\
\ref{Sec:FMNMOR}), or polarization (Sec.\ \ref{Sec:PolRes}) of the
light, the external field strength (Sec.\ \ref{Sec:ParaRes}), or
the rate of polarization relaxation (Sec.\ \ref{Sec:ParamRel})] is
modulated, and the component of the signal at a harmonic of the
modulation frequency is observed. Resonances are seen when the
modulation frequency is a subharmonic of one of the quantum-beat
frequencies present in the system.

\section{Quantum beats}
\label{Sec:QB}

\emph{Quantum beats}\cite{Har76,Dod78,Hac91,Ale93,Auz95} is the
general term for the time-evolution of a coherent superposition of
nondegenerate energy eigenstates at a frequency determined by the
energy splittings. In this paper, we are primarily concerned with
the evolution of a polarized ensemble of particles with a given
angular momentum that have their Zeeman components split by an
external field. For linear Zeeman splitting---the lowest-order
effect due to a uniform magnetic field---the evolution is Larmor
precession, i.e., rotation of the polarization about the magnetic
field direction. For nonlinear splittings, such as quadratic Stark
shifts, the polarization evolution is more complex. The state of
the ensemble is described by the density matrix, which evolves
according to the Liouville equation. While this all that is
necessary for a theoretical description of the system, physical
insight can often be gained by decomposing the density matrix into
polarization moments having the symmetries of the spherical
harmonics. In quantum beats due to nonlinear energy splittings the
relative magnitudes of the various polarization moments change
with time, a process sometimes known as
\emph{alignment-to-orientation conversion} (see Refs.\
\onlinecite{Bud2002RMP,Auz95} and references therein). A pictorial
illustration of the polarization state can be obtained by plotting
its angular momentum probability distribution. The polarization
moment decomposition and angular momentum probability distribution
not only aid physical intuition, but they are themselves complete
descriptions of the ensemble state, and can be used in some cases
to simplify calculations. For a more detailed discussion, see
Appendix \ref{Sec:Theory}.

As an example, we consider the nonlinear Zeeman shifts that occur
when atoms with hyperfine structure are subjected to a
sufficiently large magnetic field. For states with total electron
angular momentum $J=1/2$, such as in the alkali atoms, the shifted
frequency $\omega_m$ of each Zeeman sublevel with spin projection
$m$ along the magnetic field direction is given by the Breit-Rabi
formula\cite{Ale93}
\begin{equation}\label{Eqn:BreitRabi}
    \frac{\omega_m}{2\pi}
    =-\frac{\Delta}{2\prn{2I+1}}-g_I\mu m B
        \pm\frac{\Delta}{2}\prn{1+\frac{4m\xi}{2I+1}+\xi^2}^{1/2},
\end{equation}
where $\xi=(g_J+g_I)\mu B/\Delta$, $g_J$ and $g_I$ are the
electronic and nuclear Land\'{e} factors, respectively, $B$ is the
magnetic field strength, $\Gm$ is the Bohr magneton, $\Delta$ is
the hyperfine-structure interval, $I$ is the nuclear spin, and the
$\pm$ sign refers to the upper and the lower hyperfine level,
respectively. We set $\hbar=1$ throughout this paper.

Consider an atomic sample of cesium ($I=7/2$), initially in a
stretched state ($m_z=F=4$) with respect to the $z$-axis. In the
presence of an $\uv{x}$-directed magnetic field, the energy
eigenstates are the $\ket{Fm_x}$ eigenstates of the $F_x$
operator. The stretched state along $\uv{z}$ is a superposition of
these nondegenerate eigenstates, and so quantum beats are seen in
the evolution of the system.

\begin{figure}
    \includegraphics{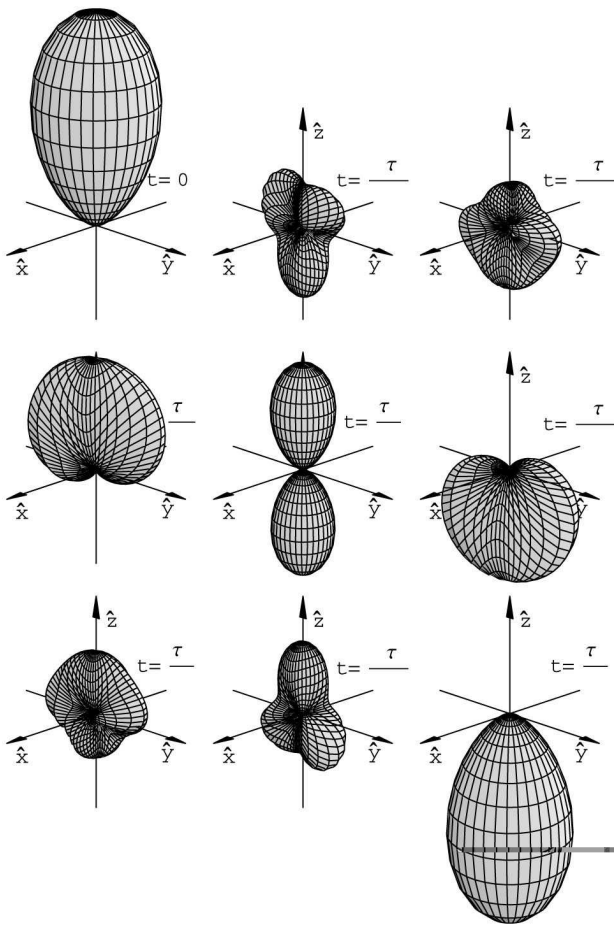}
    \caption{Quantum beats in Cs illustrated with surfaces
    representing the probability of finding the system in the
    state with maximal projection $m=F$ in a given
    direction.\cite{Auz97,Roc2001} This sequence is
    ``stroboscopic'' in the sense that the surfaces correspond to
    times chosen to have the same phase of the fast Larmor
    precession around the direction of the magnetic field
    ($\uv{x}$). From the symmetry of the plots one clearly sees
    that orientation present in the initial state collapses and
    revives in the process of the temporal evolution. Temporal
    variation of higher polarization moments give rise to
    higher-order-symmetry contributions to the probability surface
    (see also Fig.\ \ref{fig:ColRevMoments}).}
    \label{fig:ColRevProbSurf}
\end{figure}

\begin{figure}
    \includegraphics[width=3.5in]{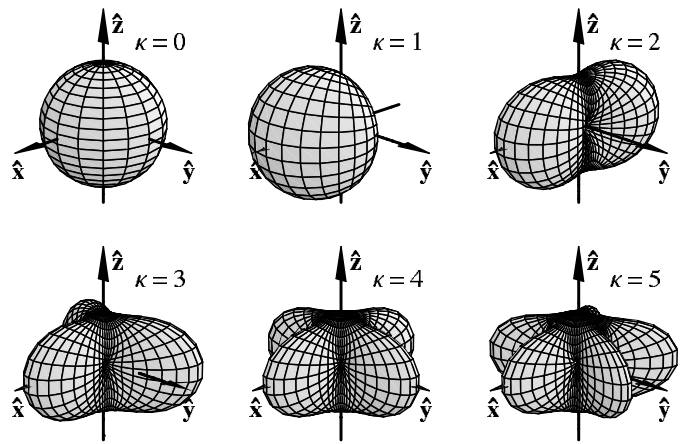}
    \caption{Angular momentum spatial distribution for states
    composed only of population ($\Gr^0_0$) and the maximum
    possible values of the components $\Gr^\Gk_{\pm\Gk}$ for a
    particular $\Gk$. $\Gk=0$: monopole moment (isotropic state
    with population only); $\Gk=1$: dipole moment (oriented
    state); $\Gk=2$: quadrupole moment (aligned state); $\Gk=3$:
    octupole moment; $\Gk=4$: hexadecapole moment; $\Gk=5$:
    triakontadipole moment.} \label{Fig:Moments}
\end{figure}

The time evolution of each eigenstate is given by
$c_m\exp\prn{-i\omega_m t}\ket{Fm}$ where $c_m$ is the initial
amplitude. For moderate field strengths such that the parameter
$\xi$ is small, the shifts deviate only slightly from linearity.
Thus, expanding Eq.\ \eqref{Eqn:BreitRabi} in powers of $\xi$, we
see that over time scales comparable to the Larmor period, the
evolution (to first order) is just Larmor precession with period
$\tau_1\simeq8(g_J\mu B)^{-1}$ (neglecting here $g_I$ compared to
$g_J$). The evolution due to the second-order quadratic shifts is
also periodic, but with a much longer period $\tau_2\simeq32\Delta
(g_J\mu B)^{-2}$. (For $B=0.5$ G, $\Gt_1\simeq6\ \Gm$s and
$\Gt_2\simeq0.3$ s.) One way to illustrate these quantum beats is
to produce graphs of the spatial distribution of angular momentum
at a given time. To do this, we plot three-dimensional closed
surfaces for which the distance from the origin in a given
direction is proportional to the probability of finding the
maximum projection of angular momentum along that
direction.\cite{Auz97,Roc2001} This plot illustrates the
symmetries of the polarization state, indicating which
polarization moments are present. (For a discussion of the angular
momentum probability distribution and the polarization moments,
see Appendix \ref{Sec:Theory}.) A collection of surfaces showing
the time-evolution of the polarization over half of a period
$\Gt_2$ of the second-order evolution is shown in Fig.\
\ref{fig:ColRevProbSurf}.
The first plot represents the initial stretched state---the
surface is literally stretched in the $\uv{z}$ direction. This
state undergoes rapid precession around $\uv{x}$ with period
$\Gt_1$. At the same time, the slower second-order evolution
results in changes in the shape of the probability surface. By
``stroboscopically'' drawing successive surfaces each at the same
phase of the fast Larmor precession (i.e., at integer multiples of
$\tau_1$), the polarization can be seen to evolve into states with
higher-order symmetry before becoming stretched along $-\uv{z}$ at
$t=\tau_2/2$. In particular, at $t=\tau_2/4$ the state is
symmetric with respect to the $x$-$y$ plane, a characteristic of
the even orders in the decomposition of the polarization state
into irreducible tensor moments $\Gr^\Gk$ (Appendix
\ref{Sec:Theory}). In general, for a multipole moment of rank
$\kappa$ with polarization transverse to the quantization axis
(such that only the components $\Gr^\Gk_q$ with $q=\pm\kappa$ are
nonzero) this moment has rotational symmetry of order $\kappa$
about that axis\cite{Auz91,Auz95} (Fig.\ \ref{Fig:Moments}).

\begin{figure}
    \includegraphics{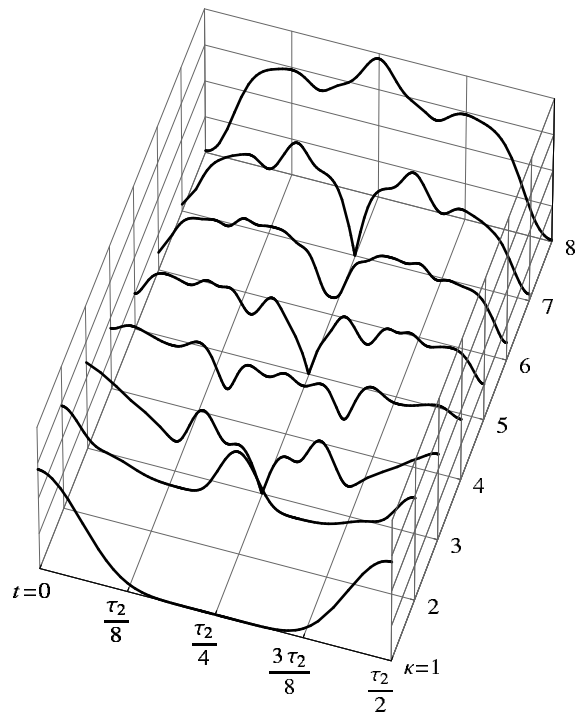}
    \caption{Temporal evolution of the norms of various
    polarization moments of ranks $\kappa$ of the $F=4$ ground
    state of Cs corresponding to the case of Figs.\
    \ref{fig:ColRevProbSurf} and \ref{fig:ColRev}. The initial
    stretched state is dominated by the lowest-order moments; at
    $t=\Gt_2/4$ the state is composed only of even-order moments.}
    \label{fig:ColRevMoments}
\end{figure}

In order to explore the decomposition into polarization moments
further, it is useful to plot the norms\cite{Roc2001} of the
polarization moments as a function of time (Fig.\
\ref{fig:ColRevMoments}).
The figure shows that initially the lowest-order moments
predominate. At $t=\tau_2/4$ the odd-order moments are zero and
the state is comprised of even-order moments only.

We can now connect these pictures of the atomic polarization state
to an experimentally observable signal, e.g., the absorption of
weak, circularly polarized light propagating along $\uv{z}$. To
find the absorption coefficient, assuming that the upper-state
hyperfine structure is not resolved, we transform to the
$\ket{J,m_J}\ket{I,m-m_J}$ basis with quantization axis along
$\uv{z}$ and sum over the transition rates for the Zeeman
sublevels.
\begin{figure}
    \includegraphics{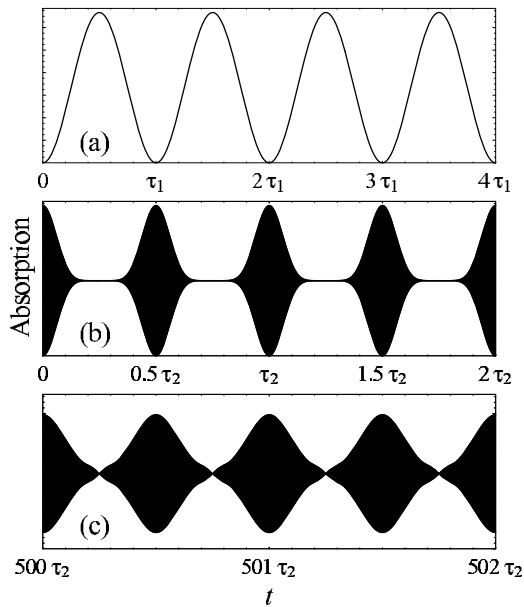}
    \caption{Collapse and revival beats arising in optically pumped Cs
    atoms due to nonlinearity of Zeeman shifts. (a) Time-dependent
    absorption of the probe light (see text) observed on a short time
    scale reveals an oscillation at the Larmor frequency. (b)
    Observation on a longer time scale reveals the characteristic
    collapse and revival (beating) behavior. Note period of
    essentially complete collapse of the oscillation pattern. (c) At
    even longer time scales, the beat pattern is modified due to
    third-order nonlinearity.} \label{fig:ColRev}
\end{figure}
At short time scales [Fig.\ \ref{fig:ColRev}(a)] we see absorption
modulated at the Larmor frequency, due to the precession of the
atomic polarization about the $x$-axis. The absorption is minimal
when the state is oriented along $\uv{z}$, and maximal when it is
oriented along $-\uv{z}$. Looking at the envelope of this
modulation at longer time scales, we see ``collapse and revival''
with period $\tau_2/2$ of the absorption oscillation amplitude
[Fig.\ \ref{fig:ColRev}(b)]. The maxima of the envelope are
associated with the stretched states shown in Fig.\
\ref{fig:ColRevProbSurf} and the minima with the states that are
symmetric with respect to the $x$-$y$ plane. This can also be seen
by comparison to Fig.\ \ref{fig:ColRevMoments}; the envelope of
the signal plotted in Fig.\ \ref{fig:ColRev}(b) is proportional to
the norm of the $\kappa=1$ moment (orientation), and does not have
any of the time-dependent behavior exhibited by the higher-order
moments in Fig.\ \ref{fig:ColRevMoments}. While it is true in
general that weak probe light is not coupled to atomic
polarization moments of rank greater than two,\cite{Dya64} the
fact that the absorption is insensitive to the $\kappa=2$ moment
(alignment) is a consequence of our assumption that the
upper-state hyperfine structure is unresolved. In Fig.\
\ref{fig:ColRev}(c) one can see the effect of the third-order
terms in the expansion of \eqref{Eqn:BreitRabi}, reducing the
contrast of the envelope function.

Collapse and revival phenomena similar to the effect described
here have been observed in nuclear precession.\cite{Maj90} In that
work, spin precession of an $I=3/2$ system, $^{201}$Hg, was
studied and the slight deviations from linearity in the Zeeman
shifts responsible for the collapse and revival beats were due to
quadrupole-interaction shifts arising from the interaction of the
atoms with the walls of a rectangular vapor cell. Collapse and
revival phenomena in molecules with large angular momenta were
considered in a tutorial paper. \cite{Auz99}


\section{Transient dynamics}
\label{Sec:PulsedPump}

In order to observe macroscopic dynamics in the magneto-optical
effects, some experimental parameter must be varied in time.
Perhaps the most conceptually straightforward technique is to
induce atomic or molecular polarization with a pulse of pump light
and then observe the transient response. Quantum beats were
originally observed in this way by detecting
fluorescence.\cite{Ale64,Dod64} We are primarily concerned here
with techniques involving probe light detection (a brief
discussion of fluorescence experiments is given in Sec.\
\ref{Sec:Fluorescence}). An early application of this method in
conjunction with probe-light polarization spectroscopy was an
experiment\cite{Lan78} with ytterbium vapor. A 5 ns pulse of
linearly polarized dye-laser light produced a coherently excited
population in the ${\rm 6\,^3P_1}$ state, whose Zeeman sublevels
were split by a magnetic field. The transmission of a subsequent
linearly polarized probe-light pulse was observed through a
crossed polarizer. Optical anisotropy in stimulated emission was
observed as polarization rotation of the probe beam. The time
dependence of the signal, due to the quantum-beat evolution
(Larmor precession) of the excited-state polarization, was
investigated indirectly by holding the probe-pulse delay time
$\tau$ fixed, and sweeping the magnetic field. Oscillations
corresponding to $\cos2\Omega_L\tau$, where $\Omega_L$ is the
Larmor frequency, were seen in the signal. The factor of two
appears because linearly polarized light induces alignment (the
$\kappa=2$ tensor moment), which has two-fold symmetry about any
axis perpendicular to the alignment axis (Sec.\ \ref{Sec:QB}).
Thus the quantum-beat frequency for this moment is $2\GO_L$; in
general, a rank $\Gk$ moment polarized as shown in Fig.\
\ref{Fig:Moments} will have quantum-beat frequency $\Gk\GO_L$ in a
$\uv{z}$-directed magnetic field.

Recently, nonlinear magneto-optical rotation with pulsed pump
light was used to study the sensitivity limits of atomic
magnetometry at very short time scales.\cite{Ger2003,Ger2004}
Observation of the time-dependence of optical rotation of a weak
probe beam allows the measurement of the magnetic field to be
corrected for the initial spin-projection uncertainty. For
measurement times $T$ short enough that non-light-induced
polarization relaxation processes can be neglected (in this case,
$T\ll(\Gg\sqrt{N})^{-1}$, where $\Gg$ is the relaxation rate and
$N$ is the number of polarized atoms), a ``quantum nondemolition
measurement''\cite{Scu97} can be performed by using far-detuned
probe light. Over this time, a sub-shotnoise measurement with
uncertainty scaling as $N^{-3/4}$ is then possible.\cite{Auz2004}
If squeezed light\cite{Scu97} is used, Heisenberg-limited scaling
of $N^{-1}$ can be obtained, limited to an even shorter
measurement time $T\ll(\Gg N)^{-1}$.

In the experiments discussed so far in this section, the
measurement times were much shorter than the polarization
relaxation time. When longer measurements are made, the amplitude
of the quantum beats will be seen to decay during the measurement,
due to mechanisms such as collisional relaxation and, for excited
states, spontaneous emission.\cite{Note:ExitedState} An often
important mechanism that has the effect of polarization relaxation
is \textit{fly-through} or \emph{transit} relaxation. This results
from polarized particles travelling out of the probe light beam
while unpolarized particles enter the beam, resulting in a
decrease in average polarization in the probe region. The
effective rate of this relaxation $\Gg_t$ can be estimated from
the average thermal velocity over the size of the relevant region.
However, this relaxation is, in general, \emph{not} described by
an exponential decay with time constant $\Gg_t$, but in many cases
can be substantially more complicated. For example, consider a
situation in which, immediately after the pump pulse, particles
are polarized in a spatial region that is larger than the probe
region. This can occur even if the pump and probe beams have
identical and overlapping profiles under conditions of nonlinear
absorption: a strong Gaussian-profiled pump beam can perform
efficient optical pumping even far away from the beam center. The
weak probe, on the other hand, acts at the intensity range of
linear absorption; it takes some time for particles in thermal
motion to reach the beam center where they undergo the probe
interaction. As a result, at the initial moments following the
pump pulse, one does not observe significant effective relaxation.
Only after a certain amount of time does fly-through relaxation
set in. This effect was predicted and measured for relaxation of
ground state K$_{2}$ molecules.\cite{Auz83} This nonexponential
relaxation kinetics can be significant for quantitative analysis
of quantum-beat signals.

In addition to the observation of transients, another way to study
quantum beat dynamics is to use resonance techniques involving
modulation of the external experimental conditions. Such
techniques are discussed in the next section.

\section{Beat resonances}
\label{Sec:BeatRes}

\subsection{Amplitude resonances}
\label{Sec:SynchPump}

In a pulsed experiment such as described in Sec.\
\ref{Sec:PulsedPump}, the quantum beats will tend to wash out as
the pulse rate is increased relative to the polarization
relaxation rate. Particles polarized during successive pulses
will, in general, beat out of phase with each other, cancelling
out the overall medium polarization. If the quantum-beat frequency
is slower than the relaxation rate, each polarized particle will
not be able to undergo an entire quantum-beat cycle before
relaxing. Thus the quantum beats will not cancel completely, and
there will be some residual steady-state polarization. This is the
effect studied in the limiting case of steady-state NMOE
experiments with cw pump light. If the quantum-beat frequency is
faster than the relaxation rate, each particle will contribute to
the average polarization over its entire quantum beat cycle, and
the macroscopic polarization of the medium will be destroyed
entirely. This is the reason that magnetometers based on
steady-state NMOE lose sensitivity for Larmor frequencies greater
than the relaxation rate. However, time-dependent macroscopic
polarization can be regained---even for high quantum beat
frequencies---if the pump light is pulsed or amplitude modulated
at a subharmonic of the quantum-beat frequency.\cite{Note:Strobe}
Polarization is produced in phase with that of particles pumped on
previous cycles. The light pulses contribute coherently to the
medium polarization, and the ensemble beats in unison. This effect
is known as \emph{synchronous optical pumping} or ``optically
driven spin precession''.\cite{Bel61a,Bel61b} This is a particular
case of a general class of phenomena known as \emph{beat
resonances},\cite{Ale63,Ale64,Ale93,Auz95} exhibited when a
parameter of the pump light or external field is modulated, as
discussed below. In NMOE experiments, such resonances are
generally observed using lock-in detection of the transmission or
polarization signal of the probe light, at a harmonic of the
modulation frequency. However, in some cases, such as when total
transmission (or fluorescence) is observed, the resonances can
also be detected in the time-averaged signal\cite{Auz90phres}.

In 1961, Bell and Bloom\cite{Bel61a} were the first to observe
beat resonances, due to quantum beats in the ground states of Rb
and Cs and in metastable He. The first ground-state quantum-beat
resonance experiments in molecules were performed with
Te$_2$\cite{fer82} and were followed by experiments with
K$_2$.\cite{Auz87,Auz90bres} Synchronous optical pumping has been
used over the years in many applications. To give just one
example, this method was employed in sensitive searches
\cite{Jac95,Rom2001PRL,Rom2001} for a possible permanent
electric-dipole moment (whose existence is only possible due to a
violation of both parity and time-reversal invariance) of
$^{199}$Hg.

\subsection{Frequency resonances}
\label{Sec:FMNMOR}

\begin{figure}
    \includegraphics[width=3.325in]{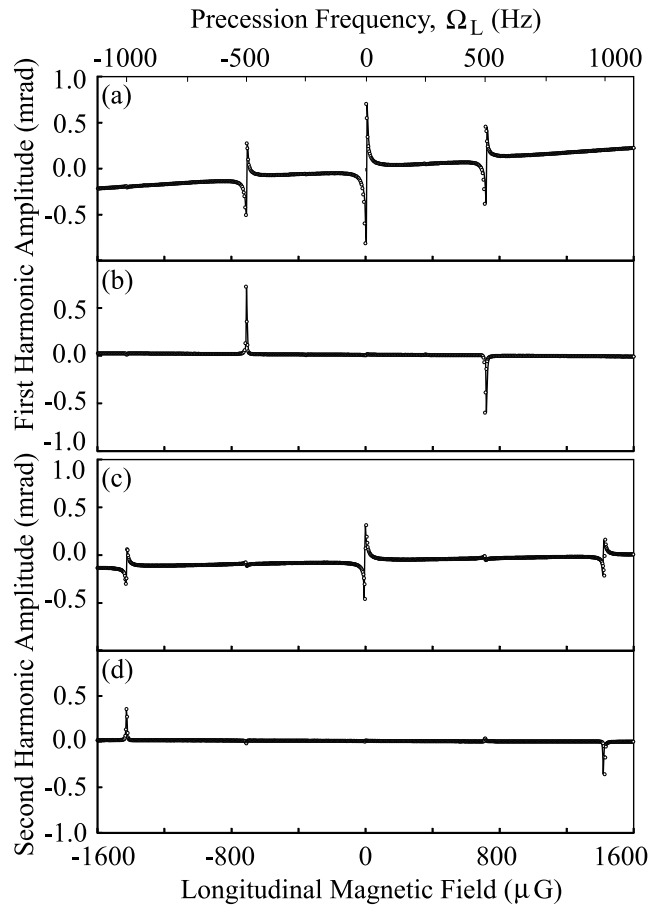}
    \caption{Signals detected at the first harmonic (a,b) and
    second harmonic (c,d) of $\Omega_m$ as a function of
    longitudinal magnetic field. This experiment employed
    buffer-gas-free, paraffin-coated vapor cells containing
    isotopically enriched $^{87}$Rb. The laser was tuned near the
    $D1$ line, laser power was 15 $\mu$W, beam diameter $\sim$2
    mm, $\Omega_m=2\pi\times1$ kHz, and the modulation amplitude
    $\Delta\omega_m = 2\pi\times220$ MHz. Traces (a,c) and (b,d)
    correspond to the in-phase and the quadrature outputs of the
    signals from the lock-in detector, respectively. The
    zero-field resonances observed in traces (a,c) are similar in
    nature to the resonances observed in static NMOE studies
    (see text). The quadrature components arise because of a
    phase difference between the ``probe'' modulation and the
    modulation of the optical properties of the atomic medium.
    (Aligned atoms produce maximum optical rotation when the
    alignment axis is at an angle of $\pi/4$ to the light
    polarization direction.) Figure from Ref.\
    \onlinecite{Bud2002FM}.} \label{Fig:FM_SvsB}
\end{figure}

Frequency (rather than amplitude) modulation of the pump light can
be used to produce an effect similar to that discussed in Sec.\
\ref{Sec:SynchPump}. Here, the optical pumping rate is modulated
as a result of its frequency dependence. One example of this is
nonlinear magneto-optical (Faraday) rotation with
frequency-modulated light (FM NMOR).
\cite{Bud2002FM,Yas2003,Mal2004} In this technique, linearly
polarized light near-resonant with an atomic transition is
directed parallel to the magnetic field. The frequency of the
light is modulated, causing the rates of optical pumping and
probing to acquire a periodic time dependence. As described in
Sec.\ \ref{Sec:SynchPump}, a resonance occurs when the
quantum-beat frequency $\kappa\Omega_L$ for a rank-$\Gk$
polarization moment equals the modulation frequency $\Omega_m$
(the lowest-order polarization moment here has $\kappa=2$). The
atomic sample is pumped into a macroscopic rotating polarized
state that causes a periodic modulation of the plane of light
polarization at the output of the medium.  The amplitude of
time-dependent optical rotation at various harmonics of $\GO_m$
can be measured with a phase-sensitive lock-in detector (Fig.\
\ref{Fig:FM_SvsB}).
Additional resonances can be observed when the quantum-beat
frequency is equal to higher harmonics of the modulation frequency
(equivalently, $\Omega_m$ is equal to subharmonics of the beat
frequency $2\Omega_L/n$, where $n$ is the harmonic order).

As discussed in Sec.\ \ref{Sec:SynchPump}, in a steady-state
nonlinear magneto-optical rotation experiment, the equilibrium
polarization of the ensemble depends on the balance of Larmor
precession with various mechanisms causing the polarization to
relax (e.g., spin-exchange collisions or wall collisions). When
the Larmor frequency is much less than the relaxation rate, the
magnitude of the optical rotation increases linearly with the
Larmor frequency. When the Larmor frequency increases,
polarization is washed out and optical rotation falls off. Such
zero-field resonances are also observed in the magnetic field
dependence of the in-phase FM NMOR signals (Fig.\
\ref{Fig:FM_SvsB}). For the zero-field resonances, $\Omega_m$ is
much faster than both $\Omega_L$ and the optical pumping rate for
the cell, so the frequency modulation does not significantly
affect the pumping process. On the other hand, as the laser
frequency is scanned through resonance, there arises a
time-dependent optical rotation, so the signal contains various
harmonics of $\Omega_m$.

The FM NMOR technique is useful for increasing the dynamic range
of NMOR-based magnetometers (Sec.\ \ref{Sec:Magnetometry}). The
beat resonances have width comparable to that of the zero-field
resonance (since the dominant relaxation mechanisms are the same),
but in principle can be centered at any desired magnetic field.

\subsection{Polarization resonances}
\label{Sec:PolRes}

In addition to beat resonances obtained by modulation of the light
amplitude and frequency, resonances due to modulation of light
polarization have also been studied (for a review of earlier work,
see Ref. \onlinecite{Ale72}). These \emph{polarization resonances}
(also called \emph{phase resonances}) are much like the other
light modulation effects discussed above. Polarization modulation
can be thought of as out-of-phase amplitude modulation of the
light polarization components, providing a conceptual link to
amplitude resonances (Sec. \ref{Sec:SynchPump}).

\begin{figure}
    \includegraphics[width=3.25in]{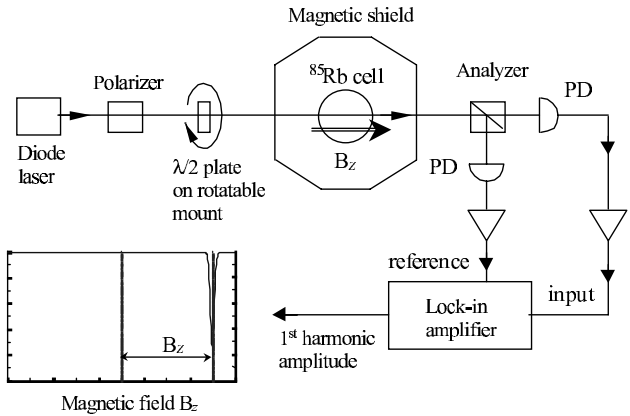}
    \caption{Experimental arrangement for the polarization
    resonance study.}
    \label{Fig:PolApp}
\end{figure}

\begin{figure}
    \includegraphics[width=3.25in]{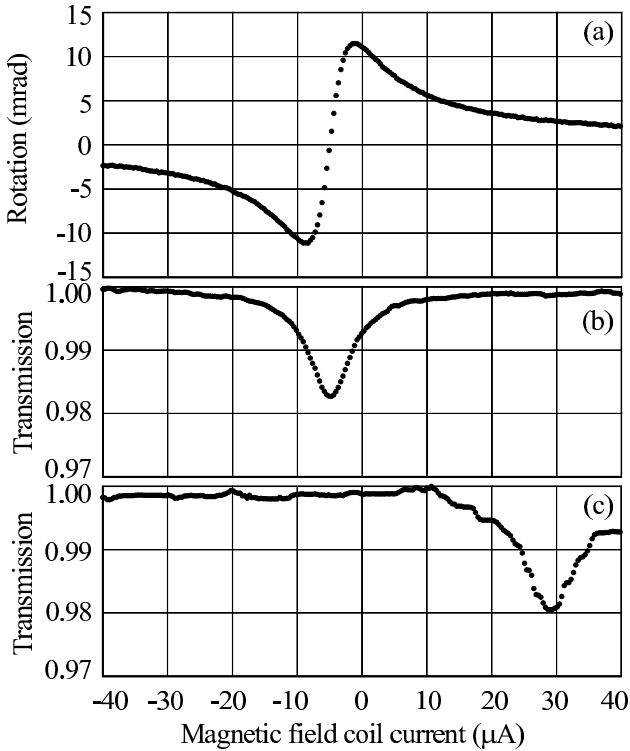}
    \caption{Faraday rotation angle (a) and transmission  (b)
    dependences on magnetic field recorded with stationary linear
    polarization of light. Trace (c) shows light transmission as a
    function of magnetic field when the linear polarization is
    rotated at $\GO_m=2\Gp\times14$ Hz. Coil current of one $\Gm$A
    corresponds to a magnetic field of approximately one $\Gm$G.}
    \label{Fig:PolData}
\end{figure}

For Zeeman beats, the light polarization, when rotated at the
Larmor frequency, is always parallel to the alignment of the
ensemble. In this case, optical pumping contributes continuously
and coherently to a rotating aligned polarization state of the
ensemble. This effect was studied in an
experiment\cite{Bud99MagProp} with Rb. A $\Gl/2$ plate on a
motor-driven rotating optical mount was used to continuously
rotate the light polarization direction at a fixed frequency;
transmission was observed with lock-in detection at this frequency
while the longitudinal magnetic field strength was swept (Fig.\
\ref{Fig:PolApp}).
The ``dark resonance'', a drop in transmission, is normally
centered at zero field when light polarization is fixed [Fig.\
\ref{Fig:PolData}(b)]. This resonance was shifted by the
polarization rotation frequency [Fig.\ \ref{Fig:PolData}(c)].
Polarization resonances occurring in molecules have been analyzed
theoretically.\cite{Auz90phres} In that work, the time-averaged
fluorescence intensity was considered as the method of detection.

\subsection{Parametric resonances}
\label{Sec:ParaRes}

As mentioned in Sec.\ \ref{Sec:SynchPump}, one way to obtain beat
resonances is to modulate the external field (e.g., the magnetic
field), and consequently the Larmor precession frequency. This
method has been used for sensitive atomic
magnetometry,\cite{Dup70} and is presently employed as a useful
general nonlinear-spectroscopic technique\cite{Fai2003} known as
\emph{parametric resonance}.

\begin{figure}
    \begin{center}
    \includegraphics[width=3.325in]{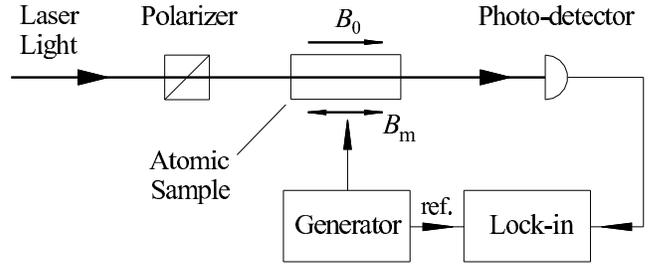}
    \caption{A schematic of an experimental arrangement for parametric
    resonance spectroscopy with magnetic-field modulation.}
    \label{Fig:ParamRes}
    \end{center}
\end{figure}

In a typical setup (Fig.\ \ref{Fig:ParamRes}), linearly polarized
resonant laser light traverses the medium, to which a magnetic
field parallel to the light-propagation direction is applied.
The magnetic field has two components---a nearly dc component
(that can be slowly scanned) and an ac component with frequency
much faster than the ground-state polarization-relaxation rate.
Transmitted light intensity is monitored with a photodetector, the
signal from which is analyzed with a lock-in amplifier referenced
to the ac modulation of the magnetic field.

As in FM NMOR (Sec.\ \ref{Sec:FMNMOR}, Fig.\ \ref{Fig:FM_SvsB}),
there are two types of resonances that are seen when the dc
magnetic field is scanned: the zero-field resonance (independent
of the ac-modulation frequency) that only appears in the in-phase
component of the signal, and the frequency-dependent resonances in
both the in-phase and quadrature components. The former arise
because for resonant light the transmission is a quadratic
function of the magnetic field, while the latter are due to an
effective modulation of the rate of polarization production. As
the ac field adds to or subtracts from the static field, it leads
to acceleration or deceleration of the Larmor precession,
respectively. Since, as discussed in Sec.\ \ref{Sec:SynchPump},
Larmor precession acts to average out the macroscopic polarization
induced by cw pump light, more overall polarization is generated
when the Larmor precession rate is the slowest. As a result, the
rate of polarization production is modulated at the ac frequency,
and so, as for amplitude or frequency resonances, a beat resonance
occurs when the modulation frequency is a subharmonic of the
quantum beat frequency.

\subsection{Parametric relaxation resonances}
\label{Sec:ParamRel}

Beat resonances can also arise when the parameter that is
modulated is the relaxation rate of the atomic polarization. Once
again, resonances occur when the frequency of the modulation
coincides with the spin-precession frequency or its subharmonic.
This effect was predicted theoretically \cite{Oku74a,Nov74} and
observed experimentally \cite{Oku74b} in metastable helium, where
relaxation rate was modulated by modulating the discharge current
in a helium cell. Note that since relaxation is independent of the
direction of the atomic spins (isotropic relaxation), this method,
in contrast to the ones discussed above, cannot produce large
overall polarizations of the sample when the Larmor precession
rate significantly exceeds the relaxation rate (see Sec.\
\ref{Sec:SynchPump}).

We also briefly mention here another related dynamic
magneto-optical effect: a sudden change in the relaxation or
optical pumping rate for a driven spin system can cause transient
spin-nutations \cite{Ale87} while the system relaxes towards its
new equilibrium state.

\subsection{Relation to coherent population trapping}
\label{Sec:CPT}

\begin{figure}
    \includegraphics[width=3.325in]{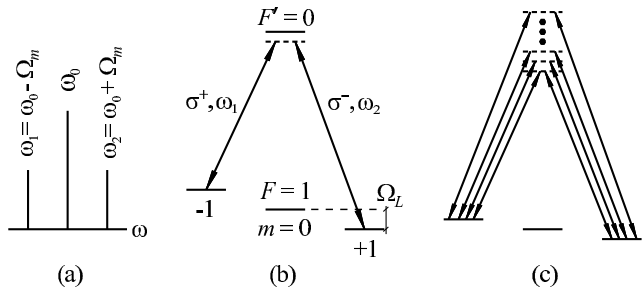}
    \caption{(a) Light frequency spectrum for light-intensity
    modulation with 50\% depth. (b) A CPT resonance at double
    modulation frequency in the case of a $F=1\rightarrow F'=0$
    transition. The lower-state Zeeman sublevels are split in a
    magnetic field applied along the quantization axis. (c) The
    CPT resonances in the case of frequency modulation with a
    large modulation index.} \label{fig:CPT_resonances}
\end{figure}

An equivalent description of the beat resonance phenomena can be
given in terms of \emph{coherent population trapping} (CPT), an
effect that is also closely related to \emph{$\Lambda$-resonances}
and \emph{mode crossing} (see Ref.\ \onlinecite{Ari96} for a
review). Indeed, harmonic modulation of the light intensity leads
to the appearance of two sidebands at frequencies shifted from the
unperturbed light frequency $\Go_0$ by the value of the modulation
frequency: $\Go_{1,2}=\Go_0\pm\GO_m$. With modulation depth less
than 100\%, the spectral component with frequency $\omega_0$ also
survives [Fig.\ \ref{fig:CPT_resonances}(a)].
A CPT resonance occurs when the frequency difference between a
pair of spectral components of the modulated light coincides with
the frequency splitting of the lower-state Zeeman sublevels, so
that light is resonant with a pair of transitions.\cite{Ari96} For
a $F=1\rightarrow F'=0$ transition [Fig.\
\ref{fig:CPT_resonances}(b)], a CPT resonance leads to transfer of
population from the ``bright'' state to the ``dark'' states that
are uncoupled to the light, and light transmission increases. For
harmonic modulation of the light intensity, there are two spectral
difference frequencies ($\GO_m,2\GO_m$), each resulting in an
observed resonance\cite{Bel61a} when the difference becomes equal
to $2\GO_L$, the splitting between the $M=1$ and $M=-1$ energy
levels.

CPT resonances can also occur when the light is frequency
modulated. In this case, the light spectrum consists of an
infinite number of sidebands with amplitudes of the $n$th sideband
given by a Bessel function $J_n(m)$ corresponding to the
modulation index $m=\Delta\omega_m/\Omega_m$, where
$\Delta\omega_m$ is the modulation depth. For example, an
experiment\cite{And2003} with cesium atoms studied the application
of the CPT effect with frequency-modulated light to atomic
magnetometry. The value of the modulation index was $m\simeq1.5$,
so only a few sidebands were prominent. Both vacuum and buffer-gas
cells were used, and the minimum observed width of the CPT
resonance was 1.4 kHz in the latter case. The width of the
resonance sets the lower bound on the magnetic fields (and,
correspondingly, the resonance frequency) for which the resonances
can be resolved.

The situation changes somewhat if the modulation index becomes
large. For example, in the nonlinear magneto-optical rotation with
frequency-modulated light experiments described in Sec.\
\ref{Sec:FMNMOR} the modulation depth is $\Delta\omega_m\simeq30$
MHz and the modulation rate is $\Omega_m\simeq 100$--1000
Hz.\cite{Bud2002FM,Yas2003,Mal2004} Thus $m\simeq10^5$ and the
pairs of $\Delta M=2$ sublevels are coupled by a very large number
of frequency-sideband pairs with comparable amplitude [Fig.\
\ref{fig:CPT_resonances}(c)]. For this reason, the description in
terms of the CPT-resonances is less intuitive here than the
synchronous-pumping picture presented in Sec.\ \ref{Sec:FMNMOR}.

\section{Applications}
\label{Sec:Apps}

\subsection{High-order polarization moments}
\label{Sec:HighOrder}

The use of Zeeman beat resonances allows one to directly create
and detect higher-order polarization moments. Because a
polarization moment with rank $\kappa$ has symmetry of order
$\kappa$ about some axis (Sec.\ \ref{Sec:QB}), a resonance in both
pumping and probing due to interaction with this moment occur when
the light-particle interaction is modulated at the frequency
$\Omega_m=\kappa\Omega_L$. A dipole transition can connect
polarization moments with ranks differing by at most two.
Consequently, multiple photon interactions are required to
generate high-rank polarization moments. For example, two
interactions are required to produce a rank $\kappa = 4$
hexadecapole moment from an initially unpolarized ($\kappa=0$)
state. There must be an equal number of photon interactions in
order to detect a signal due to the high-order multipole moment,
which will be modulated at $\kappa\Omega_L$. Thus the amplitude of
the signal due to the hexadecapole resonance scales as the fourth
power of the light intensity, as experimentally
observed\cite{Yas2003} in FM NMOR (Sec.\ \ref{Sec:FMNMOR}).

\begin{figure}
    \includegraphics[width=3.25in]{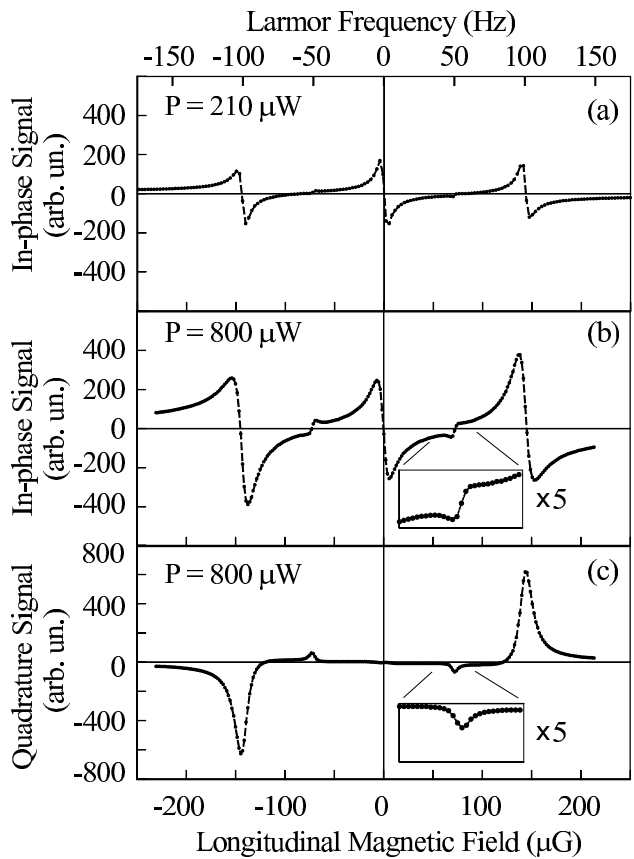}
    \caption{An example of the magnetic-field dependence of the FM
    NMOR signals showing quadrupole resonances at $B=\pm143.0$
    $\mu$G, and the hexadecapole resonances at $\pm71.5$ $\mu$G.
    Laser modulation frequency is 200 Hz, modulation amplitude is
    40 MHz peak-to-peak; the central frequency is tuned to the
    low-frequency slope of the $F=2\rightarrow F'=1$ absorption
    line. Plots (a,b) show the in-phase component of the signal at
    two different light powers; plot (c) shows the quadrature
    component. Note the increase in the relative size of the
    hexadecapole signals at the higher power. The insets show
    zooms on hexadecapole resonances. Figure from Ref.\
    \onlinecite{Yas2003}.}\label{Fig:BxDependence}
\end{figure}

Figure \ref{Fig:BxDependence} shows the magnetic-field dependence
of FM NMOR signals from a paraffin-coated cell containing
$^{87}$Rb in which the atoms are pumped and probed with a single
light beam tuned to the $D1$ transition.
At relatively low light power [Fig.\ \ref{Fig:BxDependence}(a)],
there are three prominent resonances: one at $B=0$, and two
corresponding to $2\Omega_L=\Omega_m$ (see Sec.\ \ref{Sec:FMNMOR},
Fig.\ \ref{Fig:FM_SvsB}). Much smaller signals, whose relative
amplitudes rapidly grow with light power, are seen at
$4\Omega_L=\Omega_m$, the expected positions of the hexadecapole
resonances.

Various experimental signatures of high-rank polarization moments
were previously detected with other techniques,
\cite{Sut93,Sut94,Xu97a,Xu97b,Mat2003OL} but by taking advantage
of the unique periodicity of the dynamic optical signals using
beat resonances, such high-rank moments can be directly created,
controlled, and studied.  Because of the enhanced optical
nonlinearities and different relaxation properties, higher-order
polarization moments are of particular interest for application in
many areas of quantum and nonlinear optics. In addition, when
performing high-field magnetometry in alkali atoms (Sec.\
\ref{Sec:Magnetometry}), the use of the highest-order polarization
moment supported by a given state may be advantageous, because its
evolution is free of complications due to nonlinear dependence of
the Zeeman shifts on the magnetic field.\cite{Yas2003}

\subsection{Magnetometry}
\label{Sec:Magnetometry}

The steady-state NMOE are a valuable tool for magnetometry; it has
been shown\cite{Bud2000Sens} that the sensitivity of a
magnetometer based on nonlinear Faraday rotation can reach
$\sim$$10^{-11}\,{\rm G\,Hz^{-1/2}}$. Dynamic techniques can
provide useful extensions to the steady-state methods---in
particular, they can be used to increase the magnetometer's
dynamic range by translating the narrow zero-field resonances to
higher magnetic fields. As discussed in Sec.\ \ref{Sec:FMNMOR},
steady-state NMOE lose sensitivity to magnetic fields when the
Larmor frequency is greater than the polarization relaxation rate.
If beat resonances are used for magnetic field detection, however,
the dependence of the resonance condition on the modulation
frequency can be used to tune the response of the system to a
desired magnetic field range.

While the earliest examples of beat resonance magnetometry used
amplitude\cite{Bel61a} or parametric\cite{Coh69,Dup70} resonances,
in recent years the use of light-frequency modulation has become
more common.\cite{Che95,Che96,Gil2001,Bud2002FM} This is a result
of the development and broad use of single-mode diode-laser
systems. For such lasers, frequency modulation via the diode
current and/or the cavity length controlled with a piezoelectric
transducer voltage can be simpler and more robust than either
light amplitude modulation or modulation of the applied magnetic
field. However, such frequency modulation is usually accompanied
with inevitable intensity modulation of up to
$\sim$15\%.\cite{Gil2001} The deleterious effects of parasitic
modulation and of laser-intensity noise can be avoided by
detecting optical rotation\cite{Bud2002FM} rather than
transmission.

The use of beat resonances to selectively address high-rank
polarization moments (Sec.\ \ref{Sec:HighOrder}) can also be an
advantage in magnetometry. Detection of such polarization moments
may result in increased statistical sensitivity by allowing the
use of higher light power without significant increase in the
polarization relaxation rate due to power
broadening.\cite{Yas2003}

One possible application of beat resonance magnetometers is to
low-field and remote-detection nuclear magnetic resonance and
magnetic-resonance imaging. Both parametric resonance\cite{Coh69}
and FM NMOR magnetometers\cite{Yas2004Xe} have been used to
measure the nuclear magnetization of a gaseous sample placed near
a ``probe'' Rb-vapor cell. In the latter work, the fields---due to
xenon that was polarized by spin-exchange collisions with
laser-polarized Rb--were in the nanogauss range.

\section{Related topics}

\subsection{Fluorescence detection}
\label{Sec:Fluorescence}

Another standard technique for the observation of quantum beats is
the detection, not of polarization or transmission of a probe
beam, but of spontaneous emission from an excited state. Quantum
beats in fluorescence induced by weak probe light after pulsed
excitation were experimentally observed for the first time in an
experiment\cite{Auz85,Auz86} measuring the magnetic moment of the
($v^{\prime\prime}=1,J^{\prime\prime}=73$) rovibronic level of the
ground electronic state $X^{1}\Sigma_{g}^{+}$ of K$_2$ molecules.
Many molecules have $^{1}\Sigma$ ground states. The magnetic
moment of these states is nominally zero---nonzero corrections
appear only because of mixing with other states due to
perturbation by molecular rotation.\cite{Lef2004} This makes
calculations of magnetic moments in $^{1}\Sigma$ states rather
complicated and experimental measurements of these moments become
important.

\begin{figure}
    \includegraphics[width=2.5in]{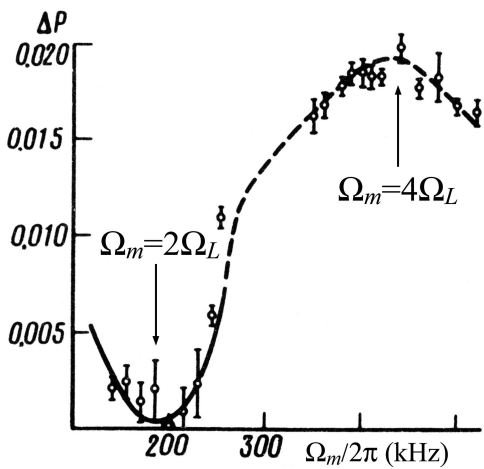}
    \caption{Ground-state beat-resonance signal measured as a
    change of the degree of polarization of laser induced
    fluorescence as a function of modulation frequency $\GO_m$ of
    excitation light. The experiment was done with K$_2$
    molecules. Two resonances, at twice (alignment) and four times
    (hexadecapole moment) the Larmor frequency $\GO_L$, were observed.
    Figure from Ref.\ \onlinecite{Auz84}.} \label{Fig:K2Res}
\end{figure}

As noted in Sec.\ \ref{Sec:HighOrder}, a single-photon dipole
interaction couples polarization moments with a difference in rank
$\Delta\kappa\le2$. Since the absorption and re-emission of probe
light is a two-photon process, moments up to rank four can be
observed in fluorescence induced by single-photon excitation.
Indeed, excited-state alignment will be directly connected by a
dipole transition to all existing ground-state moments from
population ($\kappa=0$) to the hexadecapole moment ($\kappa=4$).
This technique was used in the study of the K$_2$ ground state for
the first observation\cite{Auz84,Auz87,Auz90bres} of a beat
resonance due to the hexadecapole moment (Fig.\ \ref{Fig:K2Res}).
Relaxation dynamics\cite{Auz83} and Zeeman beats\cite{Auz86} of
multipole moments up to hexadecapole in K$_2$ molecules were also
observed using pulsed pump light and fluorescence detection.

So far, we have discussed quantum beats arising due to the
temporal evolution of the ground state. Obviously, one can also
detect quantum beats in excited-state dynamics by observing
fluorescence after pulsed excitation. This technique was used for
the first observations of quantum beats in atoms, in experiments
with the $6^{3}P_{2}$ state\cite{Ale64} of Hg and the $5^{3}P_{1}$
state\cite{Dod64} of Cd. In diatomic molecules, quantum beats from
an excited state were first observed\cite{Wal74} with iodine
dimers in the $B^{3}\Pi_{0_{u}^{+}}$ state.

Quantum-beat studies measuring fluorescence from metastable states
have been done as part of efforts to improve the present-day
limit\cite{Reg2002} on the parity- and
time-reversal-invariance-violating permanent electric-dipole
moment (EDM) of the electron.\cite{Dem2001} Close-lying states of
opposite parity that can be mixed by a static electric field are
advantageous for EDM measurements. The $J=1$ level of the first
electronic excited state of lead oxide (PbO), $a(1)\
\sbrk{^3\Sigma^+}$ is split into two opposite-parity states (the
$\Omega$-doublet\cite{Her89}) separated by only about 11 MHz. A
recent experiment\cite{Kaw2004} studied quantum beats in
fluorescence from molecules excited by a dye-laser pulse to the
$J=1$ states of $a(1)(v'=5)$. For an EDM measurement it is
necessary to measure small shifts with great accuracy; a
sensitivity of about $50\ $Hz/$\sqrt{\rm{Hz}}$ was
demonstrated,\cite{Kaw2004} consistent with shot noise in the
experiment. Precision measurements of the Land\'{e} factors of the
two components of the $\GO$-doublet were also made, laying the
groundwork for a future EDM measurement. Quantum-beat measurements
have also been done in atoms to evaluate metastable states for use
in an EDM measurement: precision tensor polarizability
measurements were done with Stark-induced quantum beats in
Sm.\cite{Roc99} Also, a pair of long-lived opposite-parity states
in Dy that are separated by only 3 MHz were studied.\cite{Bud94}
These states, components of which can be brought to crossing by
applying a weak static magnetic field, were used in a quantum-beat
technique to search for effects due to parity
violation.\cite{Ngu97}

\subsection{Polarization-noise spectroscopy}
\label{Sec:Noise}

An unpolarized medium clearly can not produce a net quantum beat
signal. However, even in an unpolarized sample the randomizing
processes that relax polarization at a rate $\Gg$ cause the medium
polarization to fluctuate around its average value of zero. Since
polarization created at a given time persists for an average time
$1/\Gg$, the polarization noise spectrum contains only components
at frequencies less than $\sim$$\Gg$ in the absence of an external
field. When a magnetic field is applied, the polarization produced
at a given time undergoes quantum beats, and so the detected
signal in probe light propagating along the magnetic field
direction is modulated at the quantum-beat frequency. Thus the
peak in the noise spectrum originally centered at zero frequency
with width $\Gg$ is shifted to the quantum-beat frequency. This
effect has been observed for Zeeman beats in an experiment on the
589 nm resonance line of Na contained in a vapor cell with buffer
gas.\cite{Ale81}

While the effect described above is classical in the sense that it
is not related to the inherent uncertainty in the polarization of
each individual particle, quantum effects can also be relevant. In
particular, the uncertainty relation between Cartesian components
of the angular momentum implies that a similar polarization noise
effect can be present not only for unpolarized samples, but also
even when there is full polarization. For example, the noise would
still be present in the same experimental geometry if the
particles were fully polarized along the magnetic field and had no
longitudinal relaxation. The width of the noise resonance would
then be given by the transverse relaxation rate.

An interesting blend of noise spectroscopy and the FM NMOR method
(Section \ref{Sec:FMNMOR}) was recently studied using nonlinear
Faraday rotation on the Rb $D$ lines.\cite{Mar2004} A balanced
polarimeter could be configured to either detect optical rotation
or ellipticity of the light transmitted through atomic vapor.
Instead of the deliberate application of laser-frequency
modulation at a given rate, the frequency noise inherent to diode
lasers was relied on. The noise power at the output of the
balanced polarimeter at a fixed frequency was observed as a
function of the magnetic field. Resonant features were seen at
$B=0$ and at the values of $B$ for which twice the Larmor
frequency coincided with the observation frequency, i.e., the
counterparts of the usual FM NMOR resonances.

\section{Conclusions and outlook}

We have discussed the various dynamic nonlinear magneto-optical
effects, including transient effects occurring when the
experimental conditions are changed suddenly, and beat resonance
effects that result from modulation of an experimental parameter.
We have attempted to bring together diverse experimental
techniques---which can involve various modulation schemes, the
``quantum'' low-$J$ limit or the ``classical'' high-$J$ limit, and
related phenomena such as coherent population trapping---and
describe them from a common viewpoint.

The dynamic effects have many applications in atomic and molecular
physics, allowing experimental methods that would be difficult to
implement using the steady-state effects. For example,
polarization moments can be separately influenced and measured,
allowing determination and exploitation of their various
properties, such as differing relaxation rates. High-sensitivity
magnetometry based on optical rotation can be performed with
arbitrarily large magnetic fields. Also, the dynamic effects
provide a robust technique for precise measurements of
energy-level splittings, which makes them an invaluable tool for
fundamental symmetry tests relating to atomic and molecular
structure and interactions.

\appendix

\section{Polarization moments and the angular momentum
probability distribution}
\label{Sec:Theory}

Here we describe in more detail the complementary descriptions of
the polarization state discussed in Sec.\ \ref{Sec:QB}. Most of
the formulas given here can be found in the literature (for
example, in Ref.\ \onlinecite{Var88}), but they can be difficult
to piece together, and, in particular, a discussion of the
connection between the quantum and classical limits is not readily
available. Thus it seems useful to gather together this
information. In this section, equation numbers of formulas found
in Ref.\ \onlinecite{Var88} are referred to in brackets.

In order to describe the polarization state of a particle, it can
be helpful to write the state as a sum of tensor operators having
the symmetries of the spherical harmonics $Y_{\Gk q}(\Gq,\Gf)$.
This multipole expansion is useful not only for understanding the
polarization symmetry (Fig.\ \ref{Fig:Moments}), but also for
reducing the complexity of the density matrix evolution equations,
especially for states with large angular momentum. In molecular
spectroscopy, one typically deals with states of much larger
angular momenta ($J\simeq100$) than for atoms. In this case, the
standard Liouville equations of motion\cite{all87} form a large
coupled system that can be difficult to solve. However, the
equations of motion for the multipole expansion coefficients can
be much simpler.\cite{Auz91} This idea was introduced by
\citet{Duc76} and later applied to the analysis of a large variety
of nonlinear magneto-optical effects in diatomic molecules (see
Ref.\ \onlinecite{Auz95} and references
therein).\cite{Note:RateEq}

We employ the polarization operators $\mc{T}^\Gk_q$, defined to be
irreducible tensors of rank $\Gk$ that satisfy the normalization
condition [Eq.\ 2.4(2)]
\begin{equation}
    \tr{\mc{T}^\Gk_q}^\dag\mc{T}^{\Gk'}_{q'}
    =\Gd_{\Gk\Gk'}\Gd_{qq'}
\end{equation}
and the phase relation [Eq.\ 2.4(3)]
\begin{equation}
    {\mc{T}^\Gk_q}^\dag=\prn{-1}^q\mc{T}^\Gk_{-q}.
\end{equation}
Matrix elements of $\mc{T}^\Gk_q$ resulting from this definition
are given by [Eq.\ 2.4.2(8)]
\begin{equation}\label{Eq:TMatrixElement}
    \bra{Jm'}\mc{T}^\Gk_q\ket{Jm}
    =\sqrt{\frac{2\Gk+1}{2J+1}}\cg{Jm\Gk q}{Jm'},
\end{equation}
where $\cg{\ldots}{\ldots}$ are the Clebsch-Gordan coefficients.
The density matrix is defined as $\Gr=\ob{\GJ\GJ^\dag}$, where
$\GJ$ is the wavefunction of one particle and the overbar denotes
the average over the ensemble. It [or any arbitrary
$(2J+1)\times(2J+1)$ Hermitian matrix] can be expanded in terms of
the polarization operators as [Eq.\ 6.1(47)]
\begin{equation}\label{Eq:PolarizationExpansion}
    \Gr
    =\sum_{\Gk=0}^{2J}
        \sum_{q=-\Gk}^\Gk
        {\Gr^\Gk_q}
        {\mc{T}^\Gk_q}^\dag,
\end{equation}
where the expansion coefficients $\Gr^\Gk_q$ are the expectation
values of $\mc{T}^\Gk_q$ [Eq.\ 6.1(48)]:
\begin{equation}
    \Gr^\Gk_q=\tr\Gr\mc{T}^\Gk_q.
\end{equation}
In terms of the density matrix elements $\Gr_{mm'}$ this gives
[Eq.\ 6.1.5.(49)]
\begin{equation}
    \Gr^\Gk_q
    =\sqrt{\frac{2\Gk+1}{2J+1}}
        \sum_{m,m'=-J}^J\cg{Jm\Gk q}{Jm'}
        \Gr_{mm'},
\end{equation}
with the inverse transformation [Eq.\ 6.1(50)]
\begin{equation}
    \Gr_{mm'}
    =\sum_{\Gk=0}^{2J}
        \sum_{q=-\Gk}^\Gk
        \sqrt{\frac{2\Gk+1}{2J+1}}
        \cg{Jm\Gk q}{Jm'}
        \Gr^\Gk_q.
\end{equation}
These relations are also commonly written in an equivalent form
using the identity
\begin{equation}
    \cg{Jm\Gk q}{Jm'}
    =\prn{-1}^{J-m}
        \sqrt{\frac{2J+1}{2\Gk+1}}
        \cg{Jm'J{-m}}{\Gk q}.
\end{equation}

To produce a visual representation of the polarization state, we
plot the probability of the maximum projection of angular momentum
along the unit vector $\uv{n}_{(\Gq,\Gf)}$, i.e., the matrix
element
$\Gr_{JJ}(\Gq,\Gf)=\bra{JJ_{(\Gq,\Gf)}}\Gr\ket{JJ_{(\Gq,\Gf)}}$,
where [Eq.\ 6.1(20)]
\begin{equation}\label{Eq:JdotnEigenfunctions}
\begin{split}
    \ket{Jm_{(\Gq,\Gf)}}
    &{}=\mc{D}(\Gf,\Gq,0)\ket{Jm}\\
    &{}=\sum_{m'}D^J_{m'm}(\Gf,\Gq,0)\ket{Jm}
\end{split}
\end{equation}
are the eigenfunctions of the $\mb{J}\cdot\uv{n}_{(\Gq,\Gf)}$
operator; the Wigner $D$-functions $D^J_{m'm}(\Ga,\Gb,\Gg)$ are
the matrix elements of the rotation operator
$\mc{D}(\Ga,\Gb,\Gg)$. Since the diagonal matrix elements of the
polarization operators are found from Eqs.\
\eqref{Eq:TMatrixElement} and \eqref{Eq:JdotnEigenfunctions} and
the properties of the $D$-functions to be [Eq.\ 6.1(27)]
\begin{equation}
    \bra{Jm_{(\Gq,\Gf)}}\mc{T}^\Gk_q\ket{Jm_{(\Gq,\Gf)}}
    =\sqrt{\frac{4\Gp}{2J+1}}\cg{Jm\Gk\ms{1}0}{Jm}Y_{\Gk q}(\Gq,\Gf),
\end{equation}
it can be seen from the expansion \eqref{Eq:PolarizationExpansion}
that the angular momentum probability distribution
\begin{equation}\label{Eq:DistributionExpansion}
    \Gr_{JJ}(\Gq,\Gf)
    =\sqrt{\frac{4\Gp}{2J+1}}
        \sum_{\Gk=0}^{2J}
        \sum_{q=-\Gk}^\Gk
        \cg{JJ\Gk\ms{1}0}{JJ}
        \Gr^\Gk_{q}
        Y^*_{\Gk q}(\Gq,\Gf)
\end{equation}
is a linear combination of spherical harmonics $Y^*_{\Gk
q}(\Gq,\Gf)$ with coefficients determined by the amplitude of the
corresponding polarization moment in the polarization state of the
ensemble. Given a probability distribution $\Gr_{JJ}(\Gq,\Gf)$,
the polarization moments $\Gr^\Gk_q$ and thus the density matrix
elements $\Gr_{mm'}$ can be recovered using the orthonormality of
the spherical harmonics, so all three are complete and equivalent
descriptions of the ensemble-averaged polarization. All three
descriptions can be useful in calculations, especially in the
large-$J$ limit, for which $\Gr_{JJ}(\Gq,\Gf)$ corresponds (apart
from a normalization factor\cite{Auz95}) to the classical
probability distribution of the angular momentum direction.

For large angular momentum, the expression for $\Gr_{JJ}(\Gq,\Gf)$
in terms of the density matrix elements $\Gr_{mm'}$ simplifies
considerably.\cite{nas81,nas99} From Eq.\
\eqref{Eq:JdotnEigenfunctions} we have
\begin{equation}
\begin{split}
    \Gr_{JJ}\prn{\Gq,\Gf}
    &{}=\bra{JJ_{(\Gq,\Gf)}}\Gr\ket{JJ_{(\Gq,\Gf)}}\\
    &{}=\sum_{m_1m_2}D^{J*}_{m_1J}D^J_{m_2J}\Gr_{m_1m_2}\\
    &{}=\sum_{m\Gm}
        D^{J*}_{m+\Gm/2,J}
        D^J_{m-\Gm/2,J}
        \Gr_{m+\Gm/2,m-\Gm/2},
\end{split}
\end{equation}
where we have used the substitution $m_{1,2}=m\pm\Gm/2$. The
$D$-functions can be evaluated for the special case of interest
here [Eq.\ 4.17(8)], giving
\begin{widetext}
\begin{equation}
    \Gr_{JJ}\prn{\Gq,\Gf}=\prn{2J}!\sum_{m\Gm}
        \frac{e^{i\Gm\Gf}
                \sbr{\cos\prn{\Gq/2}}^{2\prn{J+m}}
                \sbr{\sin\prn{\Gq/2}}^{2\prn{J-m}}
                \Gr_{m+\Gm/2,m-\Gm/2}
        }
            {\sqrt{
                \prn{J-m-\Gm/2}!
                \prn{J+m-\Gm/2}!
                \prn{J-m+\Gm/2}!
                \prn{J+m+\Gm/2}!
            }
        }\,.
\end{equation}
The factor depending on $\Gq$ can be written
\begin{equation}
    \sbr{\cos\prn{\Gq/2}}^{2\prn{J+m}}
        \sbr{\sin\prn{\Gq/2}}^{2\prn{J-m}}
    =\sbr{\frac{1}{4}\prn{1-\cos\Gq}^{1-m/J}\prn{1+\cos\Gq}^{1+m/J}}^J,
\end{equation}
and, for large $J$, has a sharp peak centered at $\cos\Gq=m/J$.
Thus as $J\rightarrow\infty$, for a given $\Gq$ only the term
$m=J\cos\Gq$ contributes to the sum, and we can write
\begin{equation}
    \Gr_{JJ}\prn{\Gq,\Gf}
    \simeq\frac{\prn{2J}!}{4^J}
        \sum_{\Gm}
        \frac{\prn{1-m/J}^{J-m}
                \prn{1+m/J}^{J+m}
                e^{i\Gm\Gf}
                \Gr_{m+\Gm/2,m-\Gm/2}
            }
            {\sqrt{\prn{J-m-\Gm/2}!
                \prn{J+m-\Gm/2}!
                \prn{J-m+\Gm/2}!\prn{J+m+\Gm/2}!}}.
\end{equation}
For physical situations of interest, we can assume that only a
limited number of polarization moments are present ($\Gk\ll2J$),
which implies that only density matrix elements
$\Gr_{m+\Gm/2,m-\Gm/2}$ with $\Gm/2\ll J$ are nonzero. For these
terms in the sum the factor with explicit dependence on $J$ can be
simplified using Stirling's approximation, $n!\simeq n^ne^{-n}$:
\begin{equation}
    \frac{\prn{2J}!}{4^J}
    \frac{
        \prn{1-m/J}^{J-m}
        \prn{1+m/J}^{J+m}
        }
    {\sqrt{
        \prn{J-m-\Gm/2}!
        \prn{J+m-\Gm/2}!
        \prn{J-m+\Gm/2}!
        \prn{J+m+\Gm/2}!
        }
    }
    \simeq\frac{\prn{2J}!}{4^J}
    \frac{
        \prn{1-m/J}^{J-m}
        \prn{1+m/J}^{J+m}
        }
    {\prn{J-m}!
    \prn{J+m}!
    }
    \simeq 1.
\end{equation}
\end{widetext}
Thus, in this limit, we have
\begin{equation}\label{Eq:rho-f}
    \Gr_{JJ}\prn{\Gq,\Gf}
    \simeq\sum_\Gm
        e^{i\Gm\Gf}\Gr_{m+\mu/2,m-\mu/2}.
\end{equation}
For example, if one uses linearly polarized light to excite a
molecular transition between states with the same $J$ in the
ground and excited state ($Q$-type transition) in the absence of a
magnetic field, only diagonal density-matrix elements will be
nonzero. We can calculate these matrix elements according to a
simple formula,\cite{Auz97}
\begin{equation}\label{Eq:f-diag}
    \Gr_{mm}=\frac{3m^2}{J(J+1)(2J+1)}.
\end{equation}
Only one summand is left in Eq.\ (\ref{Eq:rho-f}). Using
$m=J\cos\Gq$ we immediately get
\begin{equation}\label{Eq:cos^2}
    \Gr\prn{\Gq,\Gf}\simeq\frac{3\cos^2\Gq}{2J+1}.
\end{equation}

\bibliographystyle{MyJOSAB}
\bibliography{JOSAB_Dynamic,DNMOEnotes}

\begin{thebibliography}{82}
\bibitem[{Not({\natexlab{a}})}]{Note:Particle}
\bibinfo{note}{We will use the word ``particle'' to refer to either atoms or
  molecules.}
\bibitem[{Not({\natexlab{b}})}]{Note:ExitedState}
\bibinfo{note}{The quantum-beat frequency must be higher than the relaxation
  rate for quantum-beat dynamics to be clearly observed. In molecules, for
  example, the excited-state relaxation rate is often dominated by radiative
  decay and is typically $\sim$$10^{8}$ s$^{-1}$, greater than the typical
  ground-state relaxation rate ($\sim$$10^{6}$ s$^{-1}$). Thus, larger magnetic
  fields are required to observe excited-state Zeeman beats than for the
  observation of ground-state quantum beats.}
\bibitem[{Not({\natexlab{c}})}]{Note:Strobe}
\bibinfo{note}{The idea of stroboscopic matching in order to increase coupling
  between systems (light and precessing spins in this context) is also
  omnipresent in the field of nuclear magnetic resonance (NMR). For example,
  Hahn and co-workers introduced various methods of coupling spin systems with
  different gyromagnetic ratios together by matching the Larmor frequency of
  one system to the Rabi frequency of the other (the Kaplan-Hahn matching), or
  by matching the Rabi frequencies of the two systems (the Hartmann-Hahn
  matching).\cite{Sli96} A general review\cite{Hah97} of the connection between
  concepts in quantum optics and NMR was given by Hahn.}
\end{thebibliography}

\end{document}